\documentclass[fleqn,usenatbib]{mnras}

\usepackage{newtxtext,newtxmath}

\usepackage[T1]{fontenc}
\usepackage{ae,aecompl}

\usepackage{graphicx}	

\usepackage[normalem]{ulem}
\usepackage[usenames]{color}
\newcommand\corr[1]{{#1}} 
\renewcommand\sout[1]{}

\newcommand{\changaMM}{{MANGA}}
\newcommand{\be}{\begin{eqnarray}}
\newcommand{\ee}{\end{eqnarray}}

\newcommand{\msun}{\ensuremath{M_{\odot}}}
\newcommand{\rsun}{\ensuremath{R_{\odot}}}

\newcommand{\rtidal}{\ensuremath{r_{\rm T}}}
\newcommand{\Mbh}{\ensuremath{M_{\rm BH}}}
\newcommand{\Mstar}{\ensuremath{M_*}}
\newcommand{\Rstar}{\ensuremath{R_*}}
\newcommand{\rperi}{\ensuremath{r_{\rm p}}}

\newcommand{\vtidal}{\ensuremath{v_{\rm T}}}

\newcommand{\vstar}{\ensuremath{v_*}}

\title{The Effect of Impact Parameter on Tidal Disruption Events}

\author[Spaulding \& Chang]{Alexandra Spaulding$^1$\thanks{AS: spauld23@uwm.edu, PC: chang65@uwm.edu,} and Philip Chang$^{1}$
\\
$^1$ Department of Physics, University of Wisconsin-Milwaukee, 3135 North Maryland Avenue, Milwaukee, WI 53211, USA
}

\date{Accepted XXX. Received YYY; in original form ZZZ}

\pubyear{2020}
\begin{document}
\label{firstpage}
\pagerange{\pageref{firstpage}--\pageref{lastpage}}
\maketitle

\begin{abstract}
Stars that pass too close to a supermassive black hole are disrupted by the black hole's tidal gravity. Some debris is ejected while the remainder accretes into the black hole. To better study the physics of these debris, we use the moving mesh code MANGA to follow the evolution of the star from its initial encounter to its complete destruction.  By varying the impact parameter ($\beta$) of the star, we study the energy distribution of the remaining material and the fallback rate of the material into the black hole as a function of time. We show that the spread of energy in the debris and peak luminosity time ($t_{\rm peak}$) are both directly related to the impact parameter.  In particular, we find a $\beta^{1/2}$ scaling for the energy spread for $\beta=2-10$ that levels off at $\beta\gtrsim 10$.  We \corr{\sout{provide analytic arguments for the spread of energy,} discuss implication of this scaling for the} rise time of the light curve and broadness of the luminosity peak for these lower $\beta$'s. These relationships provide a possible means of inferring the impact parameters for observed TDEs.     
\end{abstract}


\section{Introduction}

A tidal disruption event (TDE) occurs when a star comes within the tidal radius of a supermassive black hole (SMBH).  Tidal forces from the SMBH overcomes the self-gravity of the star, and the star is ripped apart.  Some of the debris from the star becomes bound to the SMBH, returns to periapse and forms an accretion disk, while the remainder is unbound and goes off to infinity.  The radiation produced by the returning material forms an accretion disk and its subsequent accretion powers a highly luminous event.  
\citet{1988Natur.333..523R} showed that the material from the disrupted star returns to periapse at a rate the follows a $t^{-5/3}$ power law. Moreover, the accretion rate has peak values that are super-Eddington and produces a bright electromagnetic flare that should peak in the UV or soft X-Rays.

Over the last twenty years, a number of TDE candidates have been observed both in X-rays and transient optical facilities (for a review see \citealt{2015JHEAp...7..148K}).  These candidates have been found to broadly match these theoretical predictions.  Soft X-Ray candidates observed by ROSAT, Chandra and XMM-Newton follow a decay consistent with $t^{-5/3}$.  Optical candidates show evidence of ``slow'' ($\sim {\rm few}\times 1000\,{\rm km\,s}^{-1}$) outflows \citep[see for instance,][]{2017ApJ...842...29H,2019MNRAS.489.1463O,2019ApJ...879..119H,2019MNRAS.488.1878N,2020arXiv200602454N}, which may be indicativr of outflows from super-Eddington accretion.    


In addition to being interesting phenomenon in their own right, TDEs are also valuable probes of several areas of astrophysics.  These include black hole physics \citep{2018ApJ...852...72V}, galaxy formation, accretion physics (see for instance \citealt{2015ApJ...804...85S,2018MNRAS.478.3016W}) and stellar dynamics in galactic cores (see for instance \citealt{2020SSRv..216...35S}). TDEs can be used to signal the existence of black holes in galaxies, and the rapid increase of the fallback rate of the stellar debris introduces a new way to test our understanding of accretion physics \citep{2018ApJ...852...72V}.

The strong nonlinear hydrodynamic nature of TDEs necessitates full 3-d numerical simulations.  Initial work on TDEs have both used Lagrangian, e.g., smooth particle hydrodynamics (SPH) as well as Eulerian methods.  Both of these methodologies suffer in some aspect in simulations of TDEs.  In particular, TDEs have strong advective flows (good for SPH), shocks (good for Eulerian methods), and a large dynamic range (good for SPH, ok with AMR Eulerian).  

In the last decade, a new class of arbitrary Lagrangian-Eulerian (ALE) or moving-mesh (MM) schemes described by \citet{2010MNRAS.401..791S} has been devised as an effort to capture the best characteristics of both Lagrangian and Eulerian approaches, combining superior angular momentum conservation properties of Lagrangian schemes with superior shock capturing or Eulerian schemes.  It has been implemented in the code AREPO and has been used for many different astrophysical problems especially cosmological galaxy formation \citep[for a recent example, see for instance][]{2018MNRAS.475..676S}.  

While a number of ALE codes has been developed from \citet{2010MNRAS.401..791S}'s scheme \citep{2011ApJS..197...15D,2012ApJ...758..103G,2016A&C....16..109V,2015ApJS..216...35Y,2016ApJS..226....2D,Chang+17}, we will focus on the application of these MM schemes to the study of TDEs.  Here, \citet{2019MNRAS.487..981G} and \citet{2019MNRAS.485L.146S} have respectively used AREPO \citep{2010MNRAS.401..791S} and RICH \citep{2015ApJS..216...35Y} to study TDEs with MM methods.  Despite the obvious application to TDEs, the amount of numerical work with MM methods remains small. In this paper, we report on the new results of a study on TDEs with the moving-mesh (MM) code, \changaMM\ \citep{Chang+17,2019MNRAS.486.5809P,2020MNRAS.493.5397C}. 

Here, we focus on how the spread of energy of the debris relates to impact parameter. With this, we studied relationships between peak luminosity, peak luminosity time and impact parameter. 
Several studies have done similar research, and have found different results. \cite{2019MNRAS.485L.146S} focused on the evolution of the energy spread as a function of distance from the SMBH.  They found overall the magnitude of the energy spread to be comparable with the frozen-in approximation, which predicts the energy spread will be independent of impact parameter. Similarly, \cite{2013MNRAS.435.1809S} analytically studied TDEs and determined that the spread of energy is dominated by the freeze-in energy, although for $\beta < 5$, it is mentioned that there is slight variation in spread of energy.
\cite{2013ApJ...767...25G} simulated TDEs using FLASH \citep{2000ApJS..131..273F} and found the spread of energy initially increases
with increasing $\beta$, then a transition point of $\beta$ $\approx$ 2 is reached where the binary's gravity (or star's gravity, in their case) no longer affects the dynamics, and finally the change in energy approaches a constant. 


We have organized this paper as follows.  In \S~\ref{sec:numerical}, we describe the numerical setup, outline the included physics and initial conditions.  Much of this technology is built upon our work on common envelope evolution and, thus, we leave the detailed description to \citet{2019MNRAS.486.5809P}.  In \S~\ref{sec:results}, we describe the basic results of MM simulations of a TDE of a 1 \msun\ star and a $10^6$\msun\ BH.  Finally, we close with a discussion  in \S~\ref{sec:discussion}.  

\section{Numerical Setup}\label{sec:numerical}

We use the moving-mesh hydrodynamic solver for ChaNGa, which we call \changaMM \citep[]{Chang+17,2019MNRAS.486.5809P}, to conduct numerical simulations of TDEs.  The solver is based on the scheme described by S10, but incorporates advances in gradient estimation \citep{2016MNRAS.459.1596S} and limiters \citep{2011ApJS..197...15D}.  We also use an alternative method for constructing the Voronoi tessellation (\citealt{2009Chaos..19d1111R}; C17).  In addition, it has solvers for radiation hydrodynamics \citep{2020MNRAS.493.5397C} and general relativistic hydrodynamics \citep{2020arXiv200209613C}.  

We construct initial conditions as follows.  We first use a 1 \msun\ star evolved with MESA \citep{2011ApJS..192....3P,2013ApJS..208....4P,2015ApJS..220...15P} from the pre-main sequence to the zero age main sequence.  As in \citet{2019MNRAS.486.5809P}, we take the entropy profile and construct a star of mass $M$, whose entropy profile matches that of the original star.   
While the newly constructed star can contain a dark matter particle to model the core, which is done in common-envelope studies \citep{2019MNRAS.486.5809P}, we omit this feature in this study. 
This yields a radial profile of density and temperature that is then mapped to an unstructured particle (mesh generating point) mesh.

We construct an appropriate mesh for the star from a perturbed cubic distribution that has been periodically replicated to produce sufficient numbers of mesh-generating points.  Such a cubic distribution approximates a cubic Eulerian grid, but the points are perturbed (slightly) so that we do not suffer from degenerate faces.  We assume that each particle is representative of equal volumes, and we have rescaled their mass and temperature based on the appropriate radial position via the computed density $\rho(r)$ from MESA.  These mesh-generating points are also endowed with a radially interpolated temperature.  Outside of the star, we include a low density atmosphere of $10^{-15}$ g/cm$^3$ with temperature $10^5$ K that extends out to the total box size of $5000\rsun$, having periodic boundary conditions at its edges. The periodic boundary conditions are currently the only boundary conditions that are implemented for hydrodynamics in \changaMM.  Other boundary conditions such as inflow or outflow boundary conditions are avaiable for radiation \citep{2020MNRAS.493.5397C}.  To lower the computational cost, we follow the methodology of \citet{2019MNRAS.486.5809P} and use a mesh refinement algorithm to decrease the number of mesh-generating points in the atmosphere far from the star. We define a scale factor $\mathcal{S}(r) = (r/R_{*})^{n}$ where $R_{*}$ is the radius of the star, $r$ is the spherical radius, and $n$ is an adjustable parameter which we have set to $n = 2/3$ in this case. Starting with the same uniform replicated cubic distribution as for the star, the linear spacing between mesh-generating points is increased by $\mathcal{S}$ and their mass is increased by $\mathcal{S}^{3}$, preserving the external density.  The resulting total number of mesh-generating points is $6.9\times 10^5$ and $2.7\times 10^6$ for $10^5$ and $4\times 10^5$ mesh-generating points for the star, respectively.  In spite of our scaling of the atmospheric mesh, the majority of these points are still used to represent the low density atmosphere.
We model the SMBH as a dark matter particle with a softened gravitational length of $6.14 \times 10^11$ centimeters.  

We check that the profiles produced from the mapping from 1-D stellar evolution codes to fully 3-D hydrodynamics simulations are in reasonable hydrostatic balance both for static Newtonian stars and boosted (moving, Newtonian) stars.  In particular, we run these stars for a few dynamical times to check for stability of the profile.  While it is not expected to be in perfect hydrostatic balance due to discretization errors, these stars do not oscillate significantly.  

In our numerical experiments, we start our stars at $r=r_0$ on a zero energy orbit with a periapse of $\rperi$. 
Thus, the initial magnitude of the velocity of the star is $v_0 = \sqrt{GM/r_0} = \vtidal/\sqrt{\eta}$, where $\eta$ is the ratio between $r_0$ and $\rtidal$: $\eta = r_0/\rtidal$. To set up the components of the velocity, it is helpful to define a few dimensionless parameters.
The key length-scale in the TDE is the tidal radius, which is
\be
\rtidal = \left(\frac{\Mbh}{\Mstar}\right)^{1/3}\Rstar,
\ee
where \Mbh\ is the mass of the black hole, \Mstar\ is the mass of the star, and \Rstar\ is the radius of the star.  We define $\alpha = (\Mbh/\Mstar)^{1/3}$.  Following \citet{2013ApJ...767...25G}, we define the penetration ratio, $\beta$ as 
\be
\beta \equiv \frac{\rtidal}{\rperi},
\ee
where \rperi\ is the periapse.  Assuming a (near-) Keplerian potential, the velocity of a star on a zero energy orbit at \rtidal\ is 
\be
\vtidal = \sqrt{\frac{2G\Mbh}{\rtidal}} = \alpha \vstar,
\ee
where $G$ is Newton's constant and $\vstar = \sqrt{2G\Mstar/\Rstar} = 437\,(\Mstar/1\,\msun)^{1/2}(\Rstar/1\,R_{\odot})^{-1/2}\,{\rm km\,s}^{-1}$ is the escape velocity from the star.  This quantity turns out to be a useful quantity to write other quantities in.  For instance, the velocity at periapse is then:
\be
v_{\rm p} = \sqrt{\frac{GM}{\rperi}} = \sqrt{\beta}\vtidal = \sqrt{\beta}\alpha\vstar.
\ee
The specific angular momentum associated with a zero energy orbit is:
\be
l = v_{\rm p}\rperi = \frac{\vtidal\rtidal}{\sqrt{\beta}} = \frac{\vstar\Rstar \alpha^2}{\sqrt{\beta}}.
\ee
In the above, we have assumed that $\Mbh\gg\Mstar$. 
Thus, the initial velocity can be broken into a radial component $v_{0,\parallel}$ and a perpendicular component $v_{0,\perp}$ with values:
\be
v_{0,\perp} = \frac{l}{r_0} = \frac{\alpha\vstar}{\eta\sqrt{\beta}} \quad\textrm{and}\quad v_{0,\parallel} = \sqrt{v_0^2 - v_{0,\perp}^2} = \alpha\vstar\sqrt{\frac {\eta\beta - 1}{\eta^2\beta}},
\ee
where $\eta = r_0/\rtidal$ is the dimensionless starting radius, and $r_0$ is starting radius of the star's orbit.

We set $\Mbh=10^6\msun$, $\Mstar=1\msun$, and $\Rstar=1\,R_{\odot}$, which gives $\alpha=100$. 
We also use an adiabatic ideal gas ($\gamma = 5/3$) equation of state in these calculations, mainly for purposes of speed and its applicability at low densities.  We set the initial distance to be $\eta=10$.

\section{Results}\label{sec:results}

\begin{figure*}
    \begin{center}
    \includegraphics[width=0.8\textwidth]{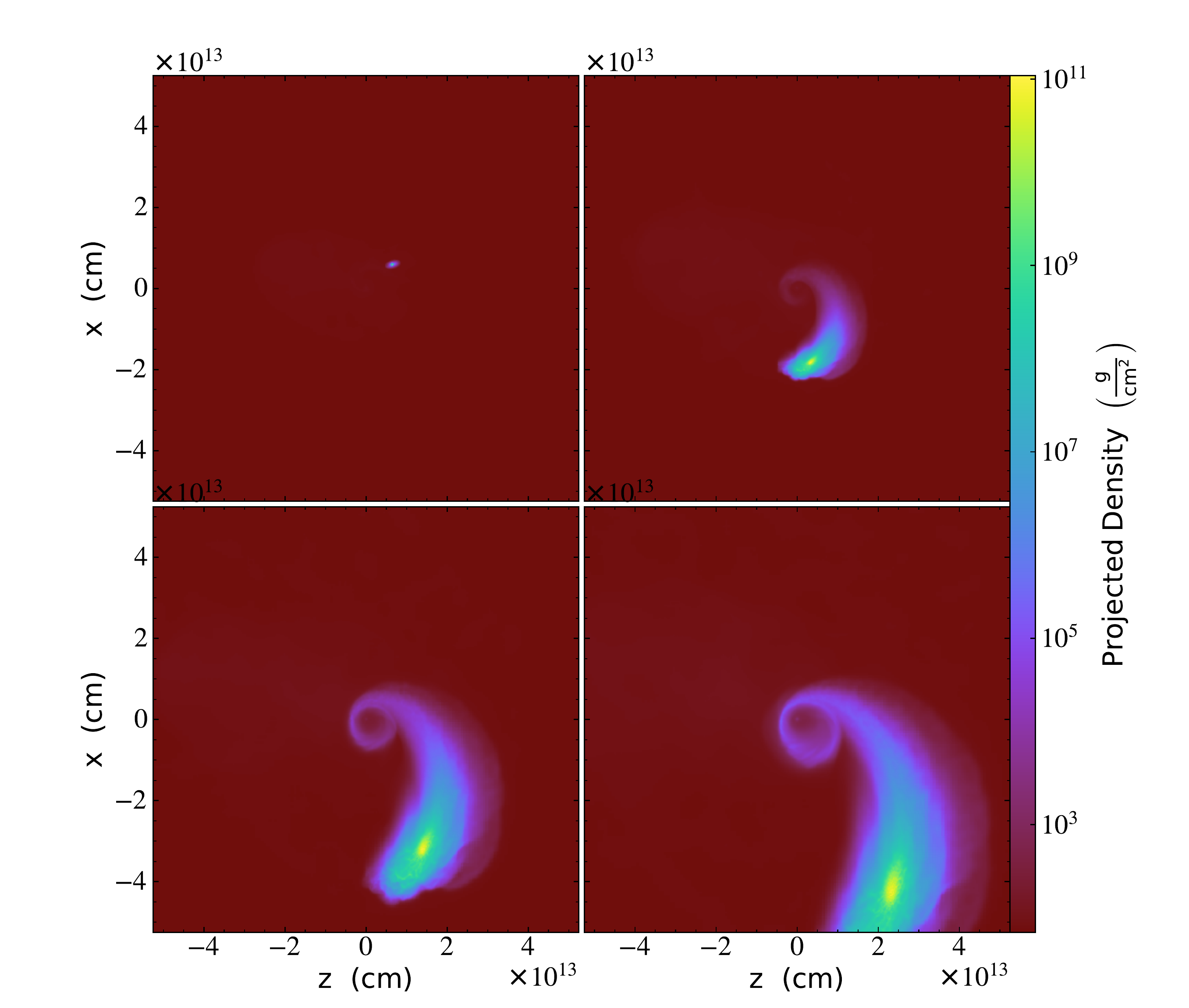}
    \end{center}
    \caption{Projected density of a 1 \msun star undergoing tidal disruption is shown from formation to 9 hours after the encounter with the black hole for $\beta=2$. Each frame is 1.8 hours apart. \label{fig:beta 2}}
\end{figure*}

\begin{figure*}
    \begin{center}
    \includegraphics[width=0.8\textwidth]{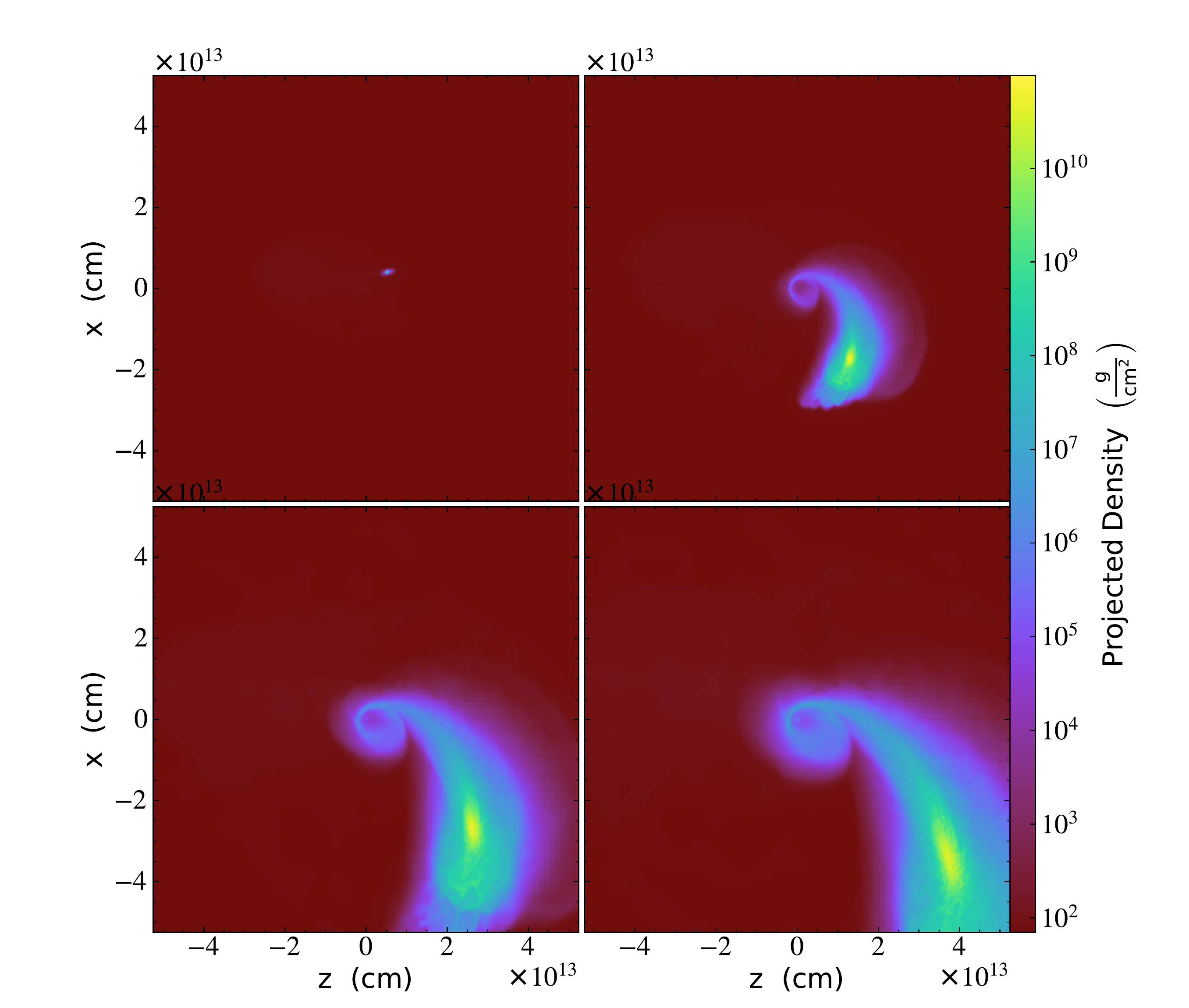}
    \end{center}
    \caption{The disruption of a star with $\beta=4$ is shown from formation to approximately 9 hours after the encounter. Each frame is 1.8 hours apart. \label{fig:beta 4}}
\end{figure*}

We vary $\beta$ between 1 and 15 in our simulations to examine the effect of the impact parameter. This parameter study is similar to previous work by \citet{2013ApJ...767...25G}, but here we use a realistic star as opposed to a polytrope and use a MM code as opposed to the Eulerian grid code, FLASH \citep{2000ApJS..131..273F}.  We use $10^5$ mesh-generating points to resolve the star for the parameter study, but confirm our result with two higher resolution studies at $4\times 10^5$ mesh-generating points.  These high resolution simulations are visualized in Figures \ref{fig:beta 2} and \ref{fig:beta 4} for $\beta = 2$ and 4, respectively.  Here we show a sequence of images for the encounter for $\beta = 2$ (Fig. \ref{fig:beta 2}) and $\beta = 4$ (Fig. \ref{fig:beta 4}).  The frames are at 3.6, 5.2, 6, and 7.8 hours from the start of the run.  The two figures show the that fallback of material appears to be much more vigorous for larger $\beta$, e.g., closer approach.  We also note that our focus is for star undergoing total disruption, which implies $\beta \gtrsim 2$ for a 1 \msun star \citep{1989IAUS..136..543P,2013ApJ...767...25G,2017A&A...600A.124M,2020arXiv200103502R}.

We have run simulations with $\beta$ varying between 1-15. However, the optimistic upper limit is $\beta = 8$ for our calculation to be valid.  In particular, we assume a Newtonian potential, but black holes are relativistic objects. The BH gravitational (Schwarzchild) radius is 
\be
   r_g = \frac{2 G\Mbh}{c^2} = 2.96\times 10^{11} \left(\frac{\Mbh}{10^6\msun}\right)\,{\rm cm}
\ee
This corresponds to a $\beta_g = \beta(R_g) = 23.5(\Mbh/10^6\msun)^{-1}\alpha$. The relativistic corrections only decline (relatively) slowly with radius, e.g., the corrections decline like $r_g/r$.  We must then choose $\beta$ such that $\beta \ll \beta_g$.  We take an optimist's view and noting that the innermost stable circular orbit is at $r_{\rm ISCO} = 3 r_g$ and set an optimistic upper limit of $\beta$ close to $\beta(r_{\rm ISCO}) \approx 8$.  Hence the results for $\beta > 8$ are for the purposes of exploring parameter space, but are not reflective of the physics.

\begin{figure}
    \includegraphics[width=0.5\textwidth]{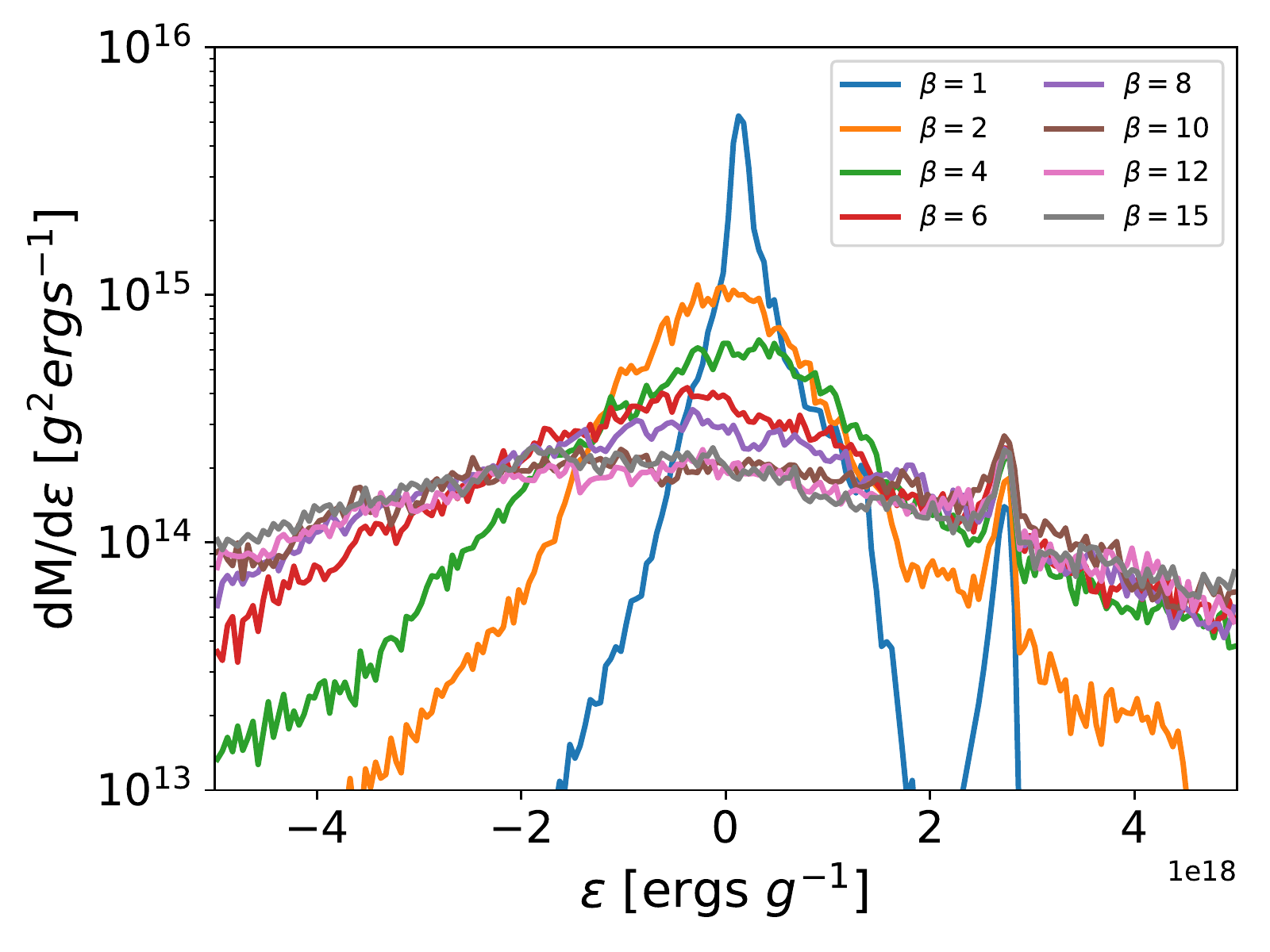}
    \caption{The mass as a function of specific energy for the $\beta=1-15$. We can see the spread in energy increases with increasing $\beta$ up to $\sim 10$. Above this value, the curve stops evolving. \label{fig:dmde}}
\end{figure}

\begin{figure}
    \includegraphics[width=0.5\textwidth]{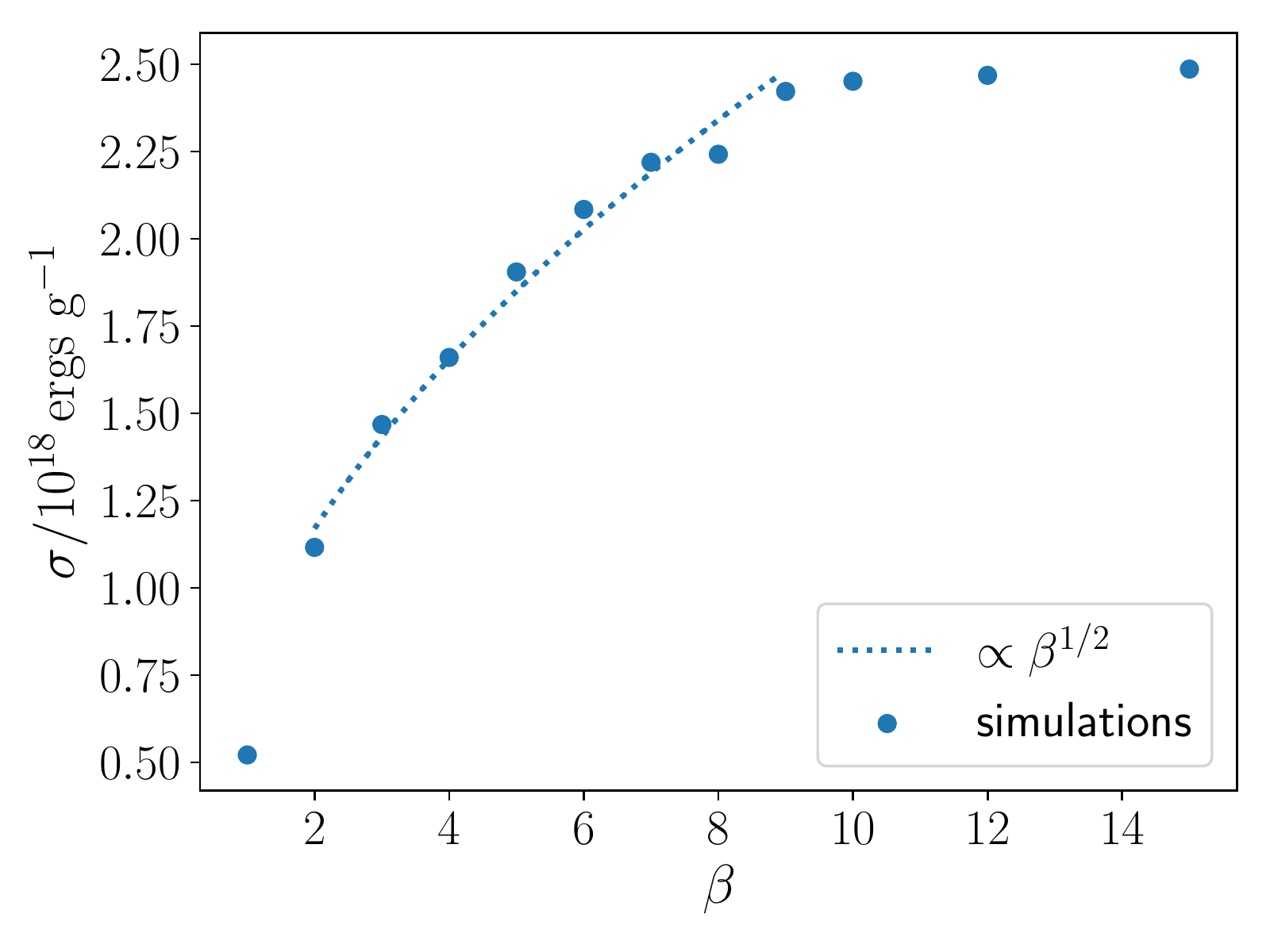}
    \caption{$\beta$ values are graphed with their corresponding $\sigma$ values. We can see a relation of $\sigma \propto \beta^{1/2}$ with the dotted line.  For  $\beta\gtrsim 10$, it transitions to the frozen-in approximation.  \label{fig:betasigma}}
\end{figure}

To understand how the fallback might vary as a function of $\beta$, we begin with the fallback rate of the debris, which is derived from Kepler's third law \citep{2019MNRAS.487..981G}:
\be
\dot{M} = \frac{dM}{dE}\frac{dE}{dt} = \frac{(2\pi G \Mbh)}{3}^{2/3}\frac{dM}{dE}t^{-5/3}. \label{eq:fallback}
\ee
From the above equation, the fallback rate is related to the distribution of energy in the material. Thus, we plot the distribution of mechanical energy in Figure \ref{fig:dmde}.  It is immediately obvious that, ignoring the peak for $\beta=1$, the spread of distribution of energy increases with increasing $\beta$. To quantify this spread, we fit the energy peak with a Gaussian of the form
\be
\frac {dM}{dE} = \frac{M_0}{\sigma} \exp\left(-\frac{(E-E_0)^2}{2\sigma^2}\right)
\ee
where $M_0$, $E_0$, and $\sigma$ are constants.  We choose a gaussian as it gives a reasonable fit over the region in energy space that contains the bulk of the mass. Here, $\sigma$ provides an estimate of the energy spread, which we list in Table \ref{tab:sigma}.  We plot $\sigma$ as a function of $\beta$ in Figure \ref{fig:betasigma} and find that it is well fit by a $\beta^{1/2}$ power law for $\beta=2-9$. We will discuss the origin of this power-law relation below. For $\beta \gtrsim 10$, the energy spread, $\sigma$, flattens, and shows a clear transition to the ``frozen-in'' approximation as discussed in \citet{2013MNRAS.435.1809S}. 

\begin{table}
	\centering
	\caption{Energy spread $\sigma$ is shown for each $\beta$ value. Our simulations show a greater jump from $\beta$=1 to higher values than previously thought.}
	\label{tab:sigma}
	\begin{tabular}{lclc} 
		\hline
		$\beta$ & $\sigma/10^{18}$ & $\beta$ & $\sigma/10^{18}$\\
		\hline
		1 & 0.52 & 7 & 2.22\\
		2 & 1.12 & 8 & 2.24\\
		3 & 1.47 & 9 & 2.42\\
		4 & 1.66 & 10 & 2.45\\
		5 & 1.91 & 12 & 2.47\\
		6 & 2.08 & 15 & 2.49\\
		\hline
	\end{tabular}
\end{table}

\begin{figure}
    \includegraphics[width=0.5\textwidth]{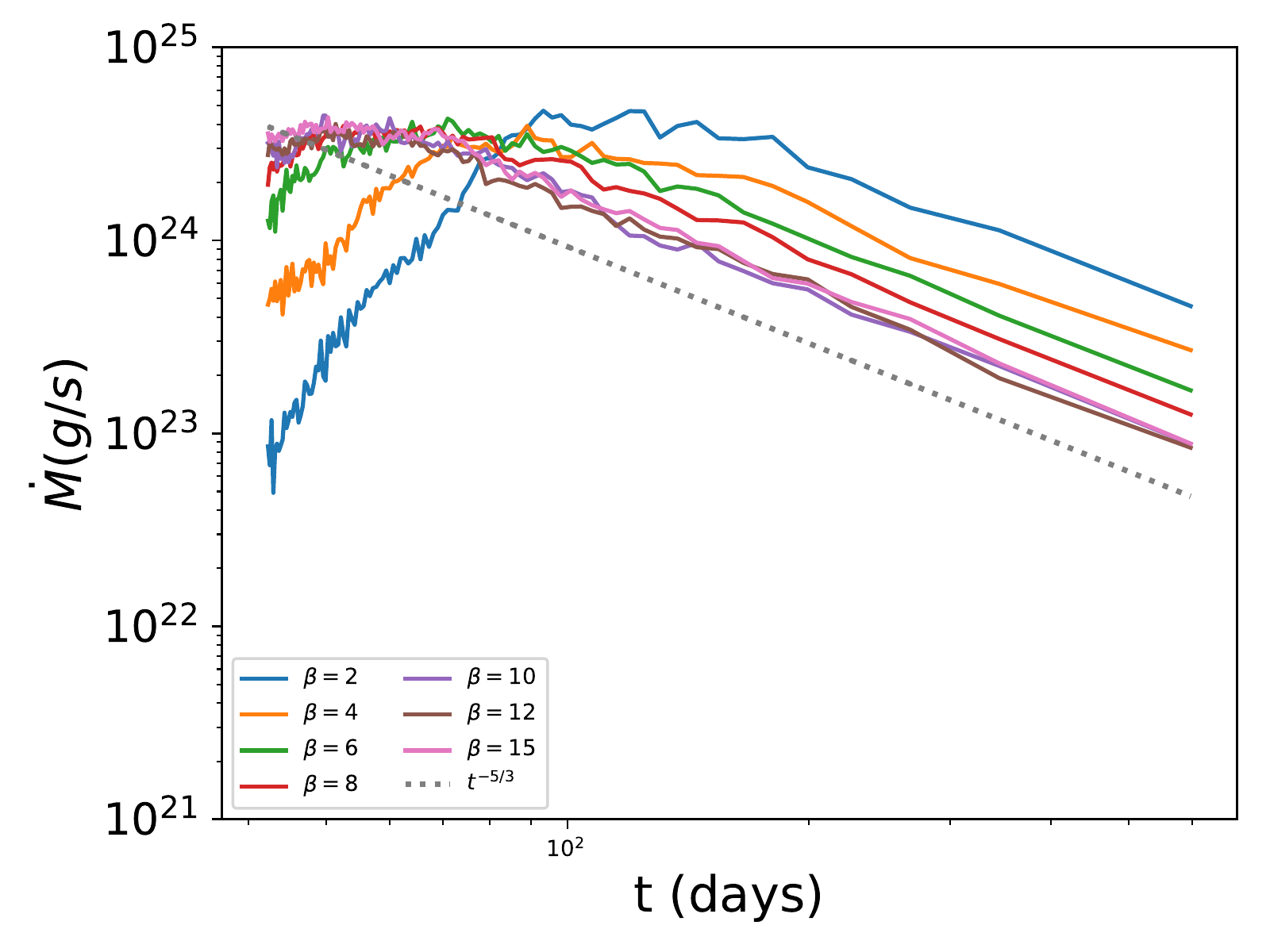}
    \caption{The fallback rate of the material into the black hole is shown to follow the theoretical rate of $t^{-5/3}$.  $\beta \lesssim 10$ are shown to have slightly different peak times for the fallback rates.  For $\beta \gtrsim 10$ , the fallback rate stops evolving with $\beta$}.\label{fig:mdot}
\end{figure}

We now use equation (\ref{eq:fallback}) to compute the fallback rate onto the black hole as a function of time and plot the result in Figure \ref{fig:mdot}. A few points are immediately noteworthy. On long timescales, $\dot{M}$ follows the theoretical fallback rate of $t^{-5/3}$ for the debris from the star.  This follows from the fact that $dM/dE$ approaches a constant value near it peaks (by definition).  Each $\beta \lesssim 10$ has a different peak luminosity time, which is about 2-4 months after the star's encounter with the SMBH. As $\beta$ is increased, the peak time is shown to slightly shift earlier.  For $\beta \gtrsim 10$, the ``frozen-in'' approximation implies that the evolution of $\dot{M}$ is fixed.

\begin{figure}
    \includegraphics[width=0.5\textwidth]{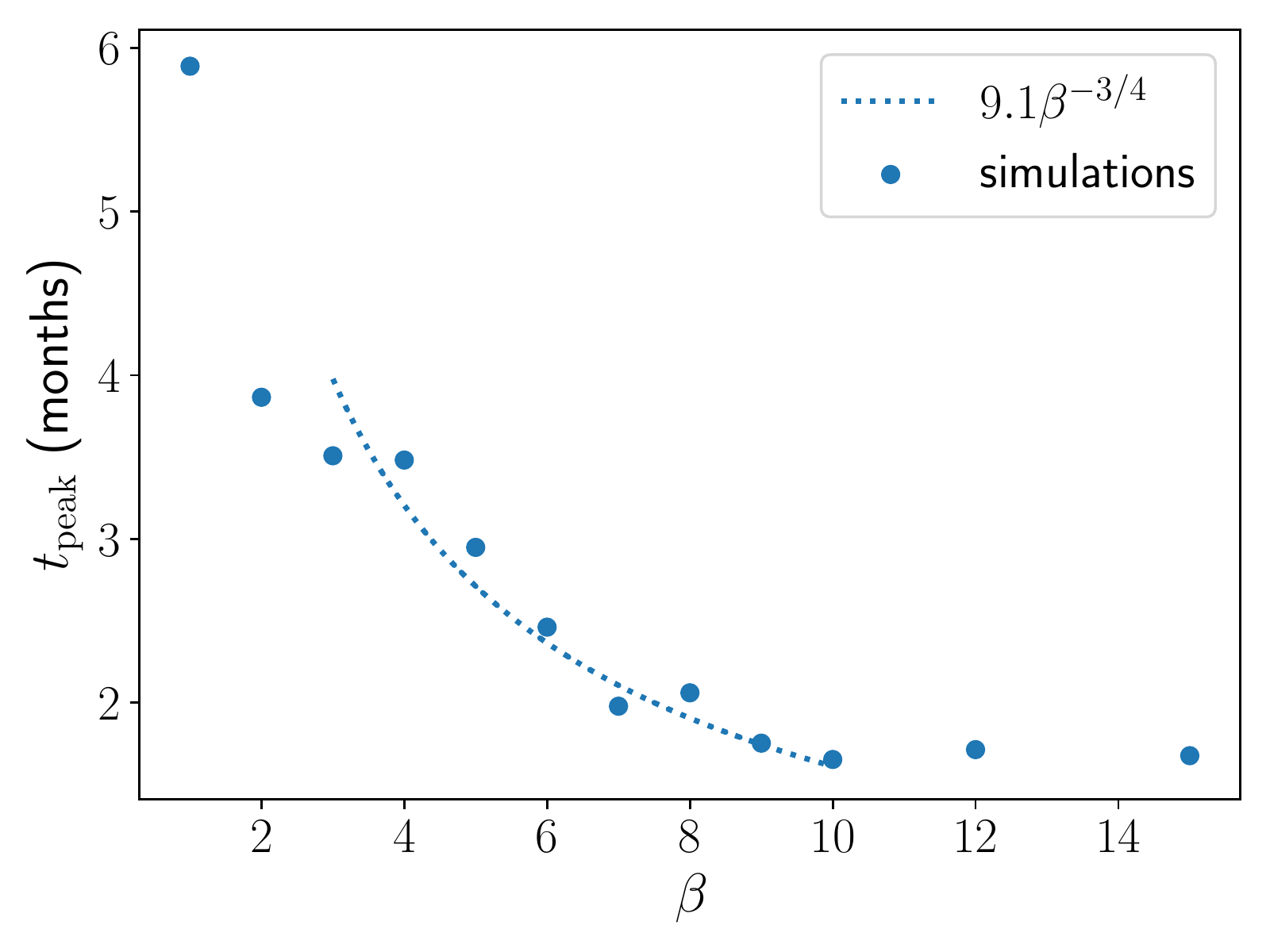}
    \caption{The peak time values corresponding to the peak fallback rate for beta values 1 through 15 are shown. The plot shows a dependence of $t_{\rm peak}\propto\beta^{-3/4}$ for $\beta \lesssim 10$. For $\beta \gtrsim 10$, this evolution stops, e.g., the "frozen-in" approximation.\label{fig:tpeak}}
\end{figure}

To find the peak time, we found the maximum $\dot{M}$ value from Figure \ref{fig:mdot} and then calculate the corresponding time. In Figure \ref{fig:tpeak} we show the relationship between peak luminosity time and $\beta$. Here we find that $t_{\rm peak}$ is progressively earlier for increasing $\beta$ up to $\approx 10$.  The earlier $t_{\rm peak}$ is in line with our previous result that the spread in $dM/dE$ scales like $\beta^{1/2}$ for $\beta \lesssim 10$.

We also plot in Figure \ref{fig:tpeak} the best fit normalization of the $t_{\rm peak}$ - $\beta$ relation to the power law $\beta^{-3/4}$.  We selected the power law based on the analysis in equation (\ref{eq:peak mdot}) below. While the fit does follow the general trend, it is not as good as the previous $\beta^{1/2}$ power law fit to $\sigma$ in Figure \ref{fig:betasigma}.
We also found no relation between $\beta$ values and the peak $\dot{M}$ values. Our results in Figure \ref{fig:mdot} show a (roughly) fixed peak $\dot{M}$ for $\beta \ge 2$.  

\begin{figure}
    \includegraphics[width=0.5\textwidth]{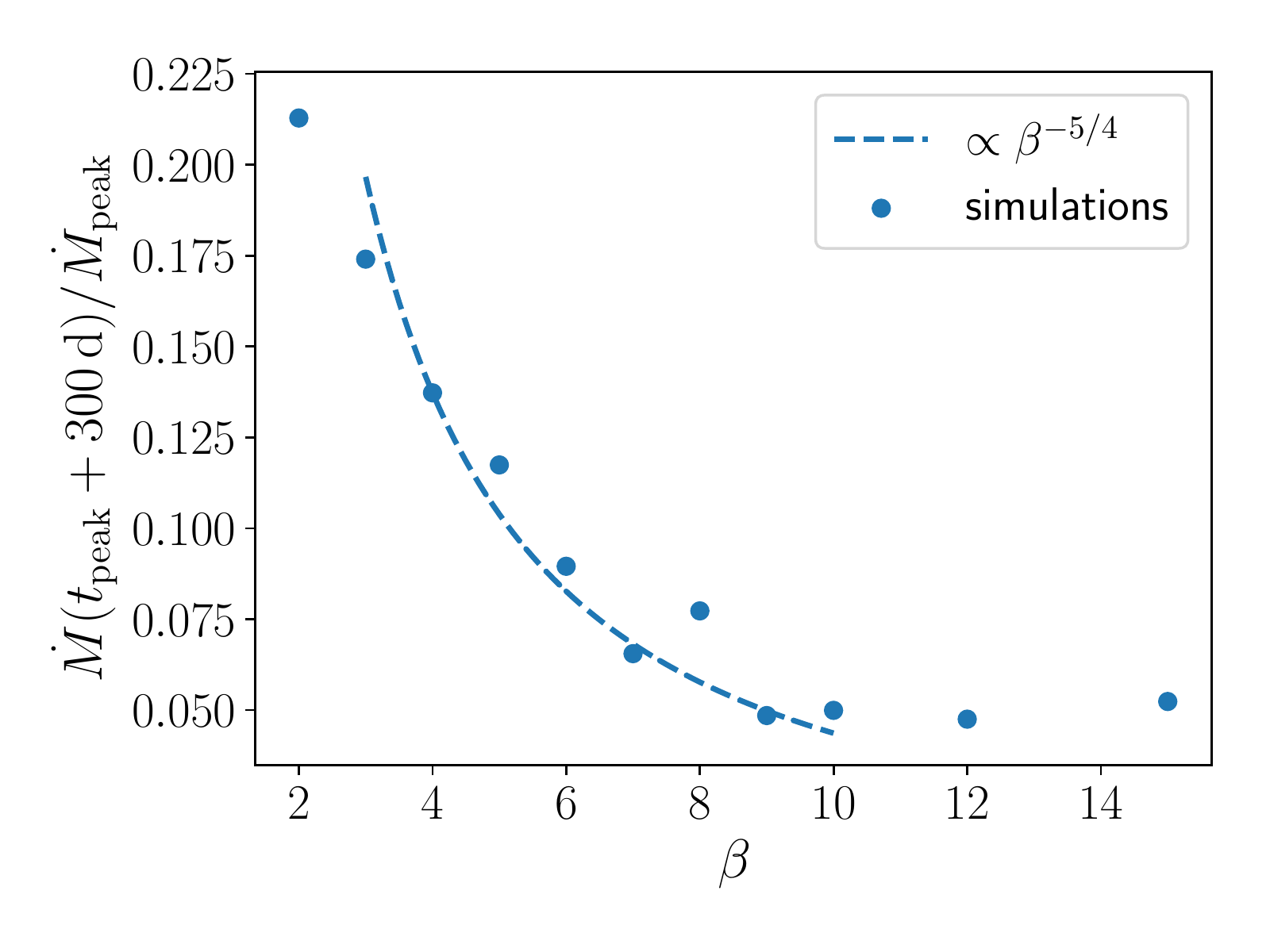}
    \caption{The fractional change of $\dot{M}$ relative to $\dot{M}_{\rm peak}$ is shown for $\beta$ values 1 through 15. The $\Delta t$ value is chosen to be 300 d. The dashed line shows a dependence of ${\dot{M}(t_{\rm peak} + \Delta t)}/{\dot{M}_{\rm peak}} \propto \beta^{-5/4}$.
    \label{fig:mdotratio}}
\end{figure}

The dependence of the peak time as a function of when the star is disrupted is difficult to pin down observationally.  However, this does have implications for the light curve.  In particular, if we assume that the mass accretion rate after the peak scales like 
\be
\dot{M}(t>t_{\rm peak}) = \dot{M}_{\rm peak}\left(\frac{t}{t_{\rm peak}}\right)^{-5/3},
\ee
then the fact that $t_{\rm peak} \propto \beta^{-3/4}$ implies that 
\be
\frac{\dot{M}(t_{\rm peak} + \Delta t)}{\dot{M}_{\rm peak}} \propto \beta^{-5/4}.\label{eq:frac mdot}
\ee
where $\Delta t$ is some time measured from $t_{\rm peak}$.  The time after $t_{\rm peak}$, $\Delta t$, should be chosen so that $\Delta t \gg t_{\rm peak}$.  In Figure \ref{fig:mdotratio}, we plot the fractional change in $\dot{M}$ relative to $\dot{M}_{\rm peak}$ for $\Delta t = 300$ d.

The dependence of $dM/dE$ on $\beta$ is not entirely clear.  Previous analytical \citep{2013MNRAS.435.1809S} and numerical \citep{2019MNRAS.485L.146S} studies have argued that $dM/dE$ should follow the frozen-in approximation, e.g., no dependence on $\beta$, for large $\beta$'s, though some allowance for a variation exists at small $\beta$.  We find that the frozen in approximation holds for $\beta \gtrsim 10$.  \corr{On the other hand, if we assume that the star stays intact to pericenter (see for instance \citealt{1988Natur.333..523R,1999ApJ...514..180U}), we find a $\beta^2$ scaling. \sout{However, we see that this is clearly not} Neither is} the case for $\beta \lesssim 10$ \corr{shown} in Figure \ref{fig:betasigma} where the spread in energy is consistent with a $\beta^{1/2}$ scaling.  This gives a factor of 2 variation over a the small dynamic range.


\corr{If we take the mild scaling of the energy spread for $\beta\approx 2-9$ as empirical, we show} two important implications.  First, let us consider the effect of this on the peak of the accretion rate.  Returning to equation (\ref{eq:fallback}), we maximize $\dot{M}$.  Setting $dM/dE$ to the Gaussian in equation (\ref{eq:fallback}) and noting that $E \propto t^{-2/3}$, we find (ignoring numerical factors):
\be \label{eq:peak mdot}
\ddot{M} = 0 \propto 2\sigma^{-2}t^{4/3} - 5 \rightarrow t_{\rm peak} \propto \sigma^{-3/2} \propto \beta^{-3/4},
\ee
using the previous scaling, $\sigma \propto \beta^{1/2}$.  This turns out to be a reasonable fit to the behavior of the $t_{\rm peak}$ in Figure \ref{fig:tpeak} for $\beta>2$ as shown by the dashed line.  Second, this leads to a variation in the change in the accretion rate (and hence luminosity) at a fixed time after the peak as shown in equation (\ref{eq:frac mdot}) above.

\section{Discussion and Conclusions}\label{sec:discussion}

In this paper, we studied the properties of the fallback rate as a function of $\beta$.  By conducting a parameter study of $\beta$ for a 1 \msun\ star and a $10^6\msun$ SMBH, we find that the spread in energy of the debris scales like $\beta^{1/2}$.  This is important because the fallback rate of the material into the black hole is directly related to $dM/dE$.  We also find that this results in an earlier $t_{\rm peak}$ for larger $\beta$.  

Previous analytical work suggests either a $\beta^2$ dependence or no dependence of $\beta$, e.g., the ``frozen-in'' approximation. We find a slow \corr{$\beta^{1/2}$} evolution in $dM/dE$ for $\beta \lesssim 10$, but recover the ``frozen-in'' approximation for $\beta \gtrsim 10$  \corr{\sout{We also present analytic arguments for both the scaling of $dM/dE$ and $t_{\rm peak}$ with $\beta \lesssim 10$.}}  We confirm convergence by running simulations varying the number of mesh-generating points that resolve the star from $10^5$ to $4\times 10^5$.  

\citet{2013ApJ...767...25G} found a similar but not exact match to our result in $\sigma$. While their work showed $\sigma$ approaching a constant with higher $\beta$, our results still show some variation as beta increases that can be explained with the relation $\sigma \propto \beta^{1/2}$. This relationship has not been mentioned in previous work. \citet{2013ApJ...767...25G} also showed a result of $t_{\rm peak}$ decreasing with increasing $\beta$ until $\beta$ = 2, which our results support.  However, our work shows $t_{\rm peak}$ to continue to decrease while they showed a slight increase in $t_{\rm peak}$ for $\beta$ > 2. Higher $\beta$ values seem to show a power law relation between peak luminosity time and $\beta$. It is still unclear how to relate this result to the theory. Similar to us, a fixed $\dot{M}$ was noted by \citet{2013ApJ...767...25G} for $\beta \ge 2$.

\cite{2019MNRAS.485L.146S} focused on the evolution of the energy spread as a function of distance from the SMBH.  They found overall the magnitude of the energy spread to be comparable with the frozen-in approximation, which predicts the energy spread to be independent of $\beta$, and that for $\beta$=1, the energy spread is larger than for the deeply penetrating cases. Our results show that the energy spread, $\sigma \propto \beta^{1/2}$, and the transition to the ``frozen-in'' approximation for larger $\beta$'s. Finally, our work agrees with \citet{2019MNRAS.487..981G}'s observation of the spread of energy in their simulations to be different than the frozen-in approximation. 

As mentioned above, one cautionary note is that for the highest $\beta$ values, relativistic corrections begin to become important.  In particular, for $\beta =8$, $r_p = 12.5\,\rsun \approx 3 r_g$, where $r_g = 2GM_{\rm BH}/c^2 = 4.2 (M_{\rm BH}/10^6 \msun)\rsun$.  At these distances, relativistic effects will begin to affect the qualitative nature of these events, e.g., direct capture is a possibility.  The transition to the ``frozen-in'' approximation occurs above this value raising the question of the relevance of this approximation for the system parameter studied in this problem.  However, we should note that for $\Mbh \sim 10^5\msun$, the ISCO is at $\beta \approx 37$, so that it is likely that the ``frozen-in'' approximation is recovered in this system.

By determining a relationship between observed TDE properties such as $t_{\rm peak}$, we can get closer to inferring $\beta$ for an observed TDE. Although our results show a fairly weak relationship between $t_{\rm peak}$ and $\beta$, it is not weak observationally. Our relationship shows differences of months in peak luminosity time between $\beta$ values, and this is a measurement we can clearly make using observations. After measuring $t_{\rm peak}$ we can figure out the evolution of $\dot M$, and this gives us two constraints on the same piece of information. From $\dot M$, we can infer stellar mass. If we take a main sequence star we know that the stellar structure does not change significantly, and with these pieces of information we can infer the $\beta$ value of an observed TDE. 

Observational searches for tidal disruption events have generally taken two paths, studies looking for X-ray transients and optical searches for supernova-like events \citep{2020SSRv..216...39M}. Modeling work such as \citet{2020arXiv200600803M} could benefit from our results. By modeling the extremely luminous transient ASASSN-15lh, they showed that a relativistic thin disc model represents the TDE observations well. However, they use a ad-hoc source term in mass that is guassian in time.  More sophisticated mass inflows from simulations such as ours can help put mutual constraint on both the mass inflow to periapse and the emission model. 


\citet{2020ApJ...894L..10H} showed that more luminous TDEs fade more slowly, but within each luminosity, there exists a bit of scatter in the rate of decline. Our results may help explain some of this scatter as we find that for the same events, the penetration parameter, $\beta$, can give a later peak and hence a broader peak. Our work could also improve studies such as these in determining impact parameter.  For instance, long followup campaigns such as the work of \citet{2020arXiv200313693H} on the TDE event ASASSN-18pg, can use the fading luminosity to constrain $\beta$.  

Care must be taken with applying these results to observations directly. First, additional physics such as black hole spin can significantly perturb the orbital trajectories of the disrupted stars and shift in time when the first self intersection of the debris may occur \citep{2015ApJ...809..166G,2019GReGr..51...30S,2019MNRAS.487.4790G}.  Second, while the mass flow rate back to pericenter can be computed, converting this to an observed luminosity is not trivial.  In particular, it remains an open question if accretion or shocks (see for instance \citealt{2015ApJ...806..164P,2015ApJ...804...85S,2020MNRAS.492..686L}) are responsible for the majority of the emission. Moreover, it is likely that outflow from disk formation or accretion changes the properties of the observed emission \citep{2015ApJ...805...83M,2018ApJ...859L..20D,2020ApJ...894....2P}.

Constraints on the distribution of $\beta$ would present an interesting probe of the dynamics near the center of galaxies \citep{2016MNRAS.455..859S}.  In particular, a flat distribution of $\beta$ or more likely a flat distribution in initial angular momentum coupled with the measure rates of TDEs would suggest that the loss-cones are full.  On the other hand, distributions that peak toward low $\beta$ would suggest empty loss cones.  In particular, \citet{2017MNRAS.468.1760W} has argued that the probability of high $\beta$ TDEs in an empty loss cone scales like $\beta^{-1}$. This would place constraints on the mass distribution in galactic centers, the degree of non-sphericity, and the general orbits of stars in the central regions (for a review see \citealt{2020SSRv..216...35S}).  

Finally, we plan future studies where we will explore the effect of radiation on the TDE debris using a recently developed radiation hydrodynamics solver \citep{2020MNRAS.493.5397C} for \changaMM.  It is expected that radiation plays a crucial role in the initial disruption \citep{2009ApJ...705..844G,2019MNRAS.482.2872Y}, self-crossing \citep{2016ApJ...830..125J}, and the final accretion disk \citep{2019ApJ...880...67J}.  With the radiation hydrodynamics solver, \changaMM\ should be particularly well suited for some of these problems.


\section*{Acknowledgements}
We thank the anonymous referee for constructive comments which greatly improve this paper. We acknowledge support by the NASA ATP program through NASA grant NNH17ZDA001N-ATP. PC acknowledges the support and hospitality of the Center for Computational Astrophysics at the Flatiron Institute where part of this work was carried out.  AS is also supported by
NASA under Award No. RFP19\_7.0 issued through the Wisconsin Space Grant Consortium and the National Space Grant College
and Fellowship Program.
We also acknowledge the Texas Advanced Computing Center (TACC) at The University of Texas at Austin for providing HPC resources that have contributed to the research results reported within this paper, \url{http://www.tacc.utexas.edu}.
We also use the yt software platform for the analysis of the data and generation of plots in this work \citep{yt}.

\section*{Data Availability}
The data underlying this article will be shared on reasonable request to the corresponding author.

\bibliographystyle{mnras}
\bibliography{references}

\begin{thebibliography}{}
\makeatletter
\relax
\def\mn@urlcharsother{\let\do\@makeother \do\$\do\&\do\#\do\^\do\_\do\%\do\~}
\def\mn@doi{\begingroup\mn@urlcharsother \@ifnextchar [ {\mn@doi@}
  {\mn@doi@[]}}
\def\mn@doi@[#1]#2{\def\@tempa{#1}\ifx\@tempa\@empty \href
  {http://dx.doi.org/#2} {doi:#2}\else \href {http://dx.doi.org/#2} {#1}\fi
  \endgroup}
\def\mn@eprint#1#2{\mn@eprint@#1:#2::\@nil}
\def\mn@eprint@arXiv#1{\href {http://arxiv.org/abs/#1} {{\tt arXiv:#1}}}
\def\mn@eprint@dblp#1{\href {http://dblp.uni-trier.de/rec/bibtex/#1.xml}
  {dblp:#1}}
\def\mn@eprint@#1:#2:#3:#4\@nil{\def\@tempa {#1}\def\@tempb {#2}\def\@tempc
  {#3}\ifx \@tempc \@empty \let \@tempc \@tempb \let \@tempb \@tempa \fi \ifx
  \@tempb \@empty \def\@tempb {arXiv}\fi \@ifundefined
  {mn@eprint@\@tempb}{\@tempb:\@tempc}{\expandafter \expandafter \csname
  mn@eprint@\@tempb\endcsname \expandafter{\@tempc}}}

\bibitem[\protect\citeauthoryear{{Chang} \& {Etienne}}{{Chang} \&
  {Etienne}}{2020}]{2020arXiv200209613C}
{Chang} P.,  {Etienne} Z.,  2020, arXiv e-prints, \href
  {https://ui.adsabs.harvard.edu/abs/2020arXiv200209613C} {p. arXiv:2002.09613}

\bibitem[\protect\citeauthoryear{{Chang}, {Wadsley}  \& {Quinn}}{{Chang}
  et~al.}{2017}]{Chang+17}
{Chang} P.,  {Wadsley} J.,   {Quinn} T.~R.,  2017, \mn@doi [\mnras]
  {10.1093/mnras/stx1809}, \href
  {http://adsabs.harvard.edu/abs/2017MNRAS.471.3577C} {471, 3577}

\bibitem[\protect\citeauthoryear{{Chang}, {Davis}  \& {Jiang}}{{Chang}
  et~al.}{2020}]{2020MNRAS.493.5397C}
{Chang} P.,  {Davis} S.~W.,   {Jiang} Y.-F.,  2020, \mn@doi [\mnras]
  {10.1093/mnras/staa573}, \href
  {https://ui.adsabs.harvard.edu/abs/2020MNRAS.493.5397C} {493, 5397}

\bibitem[\protect\citeauthoryear{{Dai}, {McKinney}, {Roth}, {Ramirez-Ruiz}  \&
  {Miller}}{{Dai} et~al.}{2018}]{2018ApJ...859L..20D}
{Dai} L.,  {McKinney} J.~C.,  {Roth} N.,  {Ramirez-Ruiz} E.,   {Miller} M.~C.,
  2018, \mn@doi [\apjl] {10.3847/2041-8213/aab429}, \href
  {https://ui.adsabs.harvard.edu/abs/2018ApJ...859L..20D} {859, L20}

\bibitem[\protect\citeauthoryear{{Duffell}}{{Duffell}}{2016}]{2016ApJS..226....2D}
{Duffell} P.~C.,  2016, \mn@doi [\apjs] {10.3847/0067-0049/226/1/2}, \href
  {http://adsabs.harvard.edu/abs/2016ApJS..226....2D} {226, 2}

\bibitem[\protect\citeauthoryear{{Duffell} \& {MacFadyen}}{{Duffell} \&
  {MacFadyen}}{2011}]{2011ApJS..197...15D}
{Duffell} P.~C.,  {MacFadyen} A.~I.,  2011, \mn@doi [\apjs]
  {10.1088/0067-0049/197/2/15}, \href
  {http://adsabs.harvard.edu/abs/2011ApJS..197...15D} {197, 15}

\bibitem[\protect\citeauthoryear{{Fryxell} et~al.,}{{Fryxell}
  et~al.}{2000}]{2000ApJS..131..273F}
{Fryxell} B.,  et~al., 2000, \mn@doi [\apjs] {10.1086/317361}, \href
  {http://adsabs.harvard.edu/abs/2000ApJS..131..273F} {131, 273}

\bibitem[\protect\citeauthoryear{{Gaburov}, {Johansen}  \& {Levin}}{{Gaburov}
  et~al.}{2012}]{2012ApJ...758..103G}
{Gaburov} E.,  {Johansen} A.,   {Levin} Y.,  2012, \mn@doi [\apj]
  {10.1088/0004-637X/758/2/103}, \href
  {http://adsabs.harvard.edu/abs/2012ApJ...758..103G} {758, 103}

\bibitem[\protect\citeauthoryear{{Gafton} \& {Rosswog}}{{Gafton} \&
  {Rosswog}}{2019}]{2019MNRAS.487.4790G}
{Gafton} E.,  {Rosswog} S.,  2019, \mn@doi [\mnras] {10.1093/mnras/stz1530},
  \href {https://ui.adsabs.harvard.edu/abs/2019MNRAS.487.4790G} {487, 4790}

\bibitem[\protect\citeauthoryear{{Goicovic}, {Springel}, {Ohlmann}  \&
  {Pakmor}}{{Goicovic} et~al.}{2019}]{2019MNRAS.487..981G}
{Goicovic} F.~G.,  {Springel} V.,  {Ohlmann} S.~T.,   {Pakmor} R.,  2019,
  \mn@doi [\mnras] {10.1093/mnras/stz1368}, \href
  {https://ui.adsabs.harvard.edu/abs/2019MNRAS.487..981G} {487, 981}

\bibitem[\protect\citeauthoryear{{Guillochon} \& {Ramirez-Ruiz}}{{Guillochon}
  \& {Ramirez-Ruiz}}{2013}]{2013ApJ...767...25G}
{Guillochon} J.,  {Ramirez-Ruiz} E.,  2013, \mn@doi [\apj]
  {10.1088/0004-637X/767/1/25}, \href
  {https://ui.adsabs.harvard.edu/abs/2013ApJ...767...25G} {767, 25}

\bibitem[\protect\citeauthoryear{{Guillochon} \& {Ramirez-Ruiz}}{{Guillochon}
  \& {Ramirez-Ruiz}}{2015}]{2015ApJ...809..166G}
{Guillochon} J.,  {Ramirez-Ruiz} E.,  2015, \mn@doi [\apj]
  {10.1088/0004-637X/809/2/166}, \href
  {https://ui.adsabs.harvard.edu/abs/2015ApJ...809..166G} {809, 166}

\bibitem[\protect\citeauthoryear{{Guillochon}, {Ramirez-Ruiz}, {Rosswog}  \&
  {Kasen}}{{Guillochon} et~al.}{2009}]{2009ApJ...705..844G}
{Guillochon} J.,  {Ramirez-Ruiz} E.,  {Rosswog} S.,   {Kasen} D.,  2009,
  \mn@doi [\apj] {10.1088/0004-637X/705/1/844}, \href
  {https://ui.adsabs.harvard.edu/abs/2009ApJ...705..844G} {705, 844}

\bibitem[\protect\citeauthoryear{{Hinkle}, {Holoien}, {Shappee}, {Auchettl},
  {Kochanek}, {Stanek}, {Payne}  \& {Thompson}}{{Hinkle}
  et~al.}{2020}]{2020ApJ...894L..10H}
{Hinkle} J.~T.,  {Holoien} T. W.~S.,  {Shappee} B.~J.,  {Auchettl} K.,
  {Kochanek} C.~S.,  {Stanek} K.~Z.,  {Payne} A.~V.,   {Thompson} T.~A.,  2020,
  \mn@doi [\apjl] {10.3847/2041-8213/ab89a2}, \href
  {https://ui.adsabs.harvard.edu/abs/2020ApJ...894L..10H} {894, L10}

\bibitem[\protect\citeauthoryear{{Holoien} et~al.,}{{Holoien}
  et~al.}{2020}]{2020arXiv200313693H}
{Holoien} T. W.~S.,  et~al., 2020, arXiv e-prints, \href
  {https://ui.adsabs.harvard.edu/abs/2020arXiv200313693H} {p. arXiv:2003.13693}

\bibitem[\protect\citeauthoryear{{Hung} et~al.,}{{Hung}
  et~al.}{2017}]{2017ApJ...842...29H}
{Hung} T.,  et~al., 2017, \mn@doi [\apj] {10.3847/1538-4357/aa7337}, \href
  {https://ui.adsabs.harvard.edu/abs/2017ApJ...842...29H} {842, 29}

\bibitem[\protect\citeauthoryear{{Hung} et~al.,}{{Hung}
  et~al.}{2019}]{2019ApJ...879..119H}
{Hung} T.,  et~al., 2019, \mn@doi [\apj] {10.3847/1538-4357/ab24de}, \href
  {https://ui.adsabs.harvard.edu/abs/2019ApJ...879..119H} {879, 119}

\bibitem[\protect\citeauthoryear{{Jiang}, {Guillochon}  \& {Loeb}}{{Jiang}
  et~al.}{2016}]{2016ApJ...830..125J}
{Jiang} Y.-F.,  {Guillochon} J.,   {Loeb} A.,  2016, \mn@doi [\apj]
  {10.3847/0004-637X/830/2/125}, \href
  {https://ui.adsabs.harvard.edu/abs/2016ApJ...830..125J} {830, 125}

\bibitem[\protect\citeauthoryear{{Jiang}, {Stone}  \& {Davis}}{{Jiang}
  et~al.}{2019}]{2019ApJ...880...67J}
{Jiang} Y.-F.,  {Stone} J.~M.,   {Davis} S.~W.,  2019, \mn@doi [\apj]
  {10.3847/1538-4357/ab29ff}, \href
  {https://ui.adsabs.harvard.edu/abs/2019ApJ...880...67J} {880, 67}

\bibitem[\protect\citeauthoryear{{Komossa}}{{Komossa}}{2015}]{2015JHEAp...7..148K}
{Komossa} S.,  2015, \mn@doi [Journal of High Energy Astrophysics]
  {10.1016/j.jheap.2015.04.006}, \href
  {https://ui.adsabs.harvard.edu/abs/2015JHEAp...7..148K} {7, 148}

\bibitem[\protect\citeauthoryear{{Lu} \& {Bonnerot}}{{Lu} \&
  {Bonnerot}}{2020}]{2020MNRAS.492..686L}
{Lu} W.,  {Bonnerot} C.,  2020, \mn@doi [\mnras] {10.1093/mnras/stz3405}, \href
  {https://ui.adsabs.harvard.edu/abs/2020MNRAS.492..686L} {492, 686}

\bibitem[\protect\citeauthoryear{{Maguire}, {Eracleous}, {Jonker}, {MacLeod}
  \& {Rosswog}}{{Maguire} et~al.}{2020}]{2020SSRv..216...39M}
{Maguire} K.,  {Eracleous} M.,  {Jonker} P.~G.,  {MacLeod} M.,   {Rosswog} S.,
  2020, \mn@doi [\ssr] {10.1007/s11214-020-00661-2}, \href
  {https://ui.adsabs.harvard.edu/abs/2020SSRv..216...39M} {216, 39}

\bibitem[\protect\citeauthoryear{{Mainetti}, {Lupi}, {Campana}, {Colpi},
  {Coughlin}, {Guillochon}  \& {Ramirez-Ruiz}}{{Mainetti}
  et~al.}{2017}]{2017A&A...600A.124M}
{Mainetti} D.,  {Lupi} A.,  {Campana} S.,  {Colpi} M.,  {Coughlin} E.~R.,
  {Guillochon} J.,   {Ramirez-Ruiz} E.,  2017, \mn@doi [\aap]
  {10.1051/0004-6361/201630092}, \href
  {https://ui.adsabs.harvard.edu/abs/2017A&A...600A.124M} {600, A124}

\bibitem[\protect\citeauthoryear{{Miller}}{{Miller}}{2015}]{2015ApJ...805...83M}
{Miller} M.~C.,  2015, \mn@doi [\apj] {10.1088/0004-637X/805/1/83}, \href
  {https://ui.adsabs.harvard.edu/abs/2015ApJ...805...83M} {805, 83}

\bibitem[\protect\citeauthoryear{{Mummery} \& {Balbus}}{{Mummery} \&
  {Balbus}}{2020}]{2020arXiv200600803M}
{Mummery} A.,  {Balbus} S.,  2020, arXiv e-prints, \href
  {https://ui.adsabs.harvard.edu/abs/2020arXiv200600803M} {p. arXiv:2006.00803}

\bibitem[\protect\citeauthoryear{{Nicholl} et~al.,}{{Nicholl}
  et~al.}{2019}]{2019MNRAS.488.1878N}
{Nicholl} M.,  et~al., 2019, \mn@doi [\mnras] {10.1093/mnras/stz1837}, \href
  {https://ui.adsabs.harvard.edu/abs/2019MNRAS.488.1878N} {488, 1878}

\bibitem[\protect\citeauthoryear{{Nicholl} et~al.,}{{Nicholl}
  et~al.}{2020}]{2020arXiv200602454N}
{Nicholl} M.,  et~al., 2020, arXiv e-prints, \href
  {https://ui.adsabs.harvard.edu/abs/2020arXiv200602454N} {p. arXiv:2006.02454}

\bibitem[\protect\citeauthoryear{{Onori} et~al.,}{{Onori}
  et~al.}{2019}]{2019MNRAS.489.1463O}
{Onori} F.,  et~al., 2019, \mn@doi [\mnras] {10.1093/mnras/stz2053}, \href
  {https://ui.adsabs.harvard.edu/abs/2019MNRAS.489.1463O} {489, 1463}

\bibitem[\protect\citeauthoryear{{Paxton}, {Bildsten}, {Dotter}, {Herwig},
  {Lesaffre}  \& {Timmes}}{{Paxton} et~al.}{2011}]{2011ApJS..192....3P}
{Paxton} B.,  {Bildsten} L.,  {Dotter} A.,  {Herwig} F.,  {Lesaffre} P.,
  {Timmes} F.,  2011, \mn@doi [\apjs] {10.1088/0067-0049/192/1/3}, \href
  {http://adsabs.harvard.edu/abs/2011ApJS..192....3P} {192, 3}

\bibitem[\protect\citeauthoryear{{Paxton} et~al.,}{{Paxton}
  et~al.}{2013}]{2013ApJS..208....4P}
{Paxton} B.,  et~al., 2013, \mn@doi [\apjs] {10.1088/0067-0049/208/1/4}, \href
  {http://adsabs.harvard.edu/abs/2013ApJS..208....4P} {208, 4}

\bibitem[\protect\citeauthoryear{{Paxton} et~al.,}{{Paxton}
  et~al.}{2015}]{2015ApJS..220...15P}
{Paxton} B.,  et~al., 2015, \mn@doi [\apjs] {10.1088/0067-0049/220/1/15}, \href
  {http://adsabs.harvard.edu/abs/2015ApJS..220...15P} {220, 15}

\bibitem[\protect\citeauthoryear{{Phinney}}{{Phinney}}{1989}]{1989IAUS..136..543P}
{Phinney} E.~S.,  1989, in {Morris} M.,  ed.,  IAU Symposium Vol. 136, The
  Center of the Galaxy. p.~543

\bibitem[\protect\citeauthoryear{{Piran}, {Svirski}, {Krolik}, {Cheng}  \&
  {Shiokawa}}{{Piran} et~al.}{2015}]{2015ApJ...806..164P}
{Piran} T.,  {Svirski} G.,  {Krolik} J.,  {Cheng} R.~M.,   {Shiokawa} H.,
  2015, \mn@doi [\apj] {10.1088/0004-637X/806/2/164}, \href
  {https://ui.adsabs.harvard.edu/abs/2015ApJ...806..164P} {806, 164}

\bibitem[\protect\citeauthoryear{{Piro} \& {Lu}}{{Piro} \&
  {Lu}}{2020}]{2020ApJ...894....2P}
{Piro} A.~L.,  {Lu} W.,  2020, \mn@doi [\apj] {10.3847/1538-4357/ab83f6}, \href
  {https://ui.adsabs.harvard.edu/abs/2020ApJ...894....2P} {894, 2}

\bibitem[\protect\citeauthoryear{{Prust} \& {Chang}}{{Prust} \&
  {Chang}}{2019}]{2019MNRAS.486.5809P}
{Prust} L.~J.,  {Chang} P.,  2019, \mn@doi [\mnras] {10.1093/mnras/stz1219},
  \href {https://ui.adsabs.harvard.edu/abs/2019MNRAS.486.5809P} {486, 5809}

\bibitem[\protect\citeauthoryear{{Rees}}{{Rees}}{1988}]{1988Natur.333..523R}
{Rees} M.~J.,  1988, \mn@doi [\nat] {10.1038/333523a0}, \href
  {https://ui.adsabs.harvard.edu/abs/1988Natur.333..523R} {333, 523}

\bibitem[\protect\citeauthoryear{{Rycroft}}{{Rycroft}}{2009}]{2009Chaos..19d1111R}
{Rycroft} C.~H.,  2009, \mn@doi [Chaos] {10.1063/1.3215722}, \href
  {http://adsabs.harvard.edu/abs/2009Chaos..19d1111R} {19, 041111}

\bibitem[\protect\citeauthoryear{{Ryu}, {Krolik}, {Piran}  \& {Noble}}{{Ryu}
  et~al.}{2020}]{2020arXiv200103502R}
{Ryu} T.,  {Krolik} J.,  {Piran} T.,   {Noble} S.~C.,  2020, arXiv e-prints,
  \href {https://ui.adsabs.harvard.edu/abs/2020arXiv200103502R} {p.
  arXiv:2001.03502}

\bibitem[\protect\citeauthoryear{{Shiokawa}, {Krolik}, {Cheng}, {Piran}  \&
  {Noble}}{{Shiokawa} et~al.}{2015}]{2015ApJ...804...85S}
{Shiokawa} H.,  {Krolik} J.~H.,  {Cheng} R.~M.,  {Piran} T.,   {Noble} S.~C.,
  2015, \mn@doi [\apj] {10.1088/0004-637X/804/2/85}, \href
  {https://ui.adsabs.harvard.edu/abs/2015ApJ...804...85S} {804, 85}

\bibitem[\protect\citeauthoryear{{Springel}}{{Springel}}{2010}]{2010MNRAS.401..791S}
{Springel} V.,  2010, \mn@doi [\mnras] {10.1111/j.1365-2966.2009.15715.x},
  \href {http://adsabs.harvard.edu/abs/2010MNRAS.401..791S} {401, 791}

\bibitem[\protect\citeauthoryear{{Springel} et~al.,}{{Springel}
  et~al.}{2018}]{2018MNRAS.475..676S}
{Springel} V.,  et~al., 2018, \mn@doi [\mnras] {10.1093/mnras/stx3304}, \href
  {https://ui.adsabs.harvard.edu/abs/2018MNRAS.475..676S} {475, 676}

\bibitem[\protect\citeauthoryear{{Steinberg}, {Yalinewich}  \&
  {Sari}}{{Steinberg} et~al.}{2016}]{2016MNRAS.459.1596S}
{Steinberg} E.,  {Yalinewich} A.,   {Sari} R.,  2016, \mn@doi [\mnras]
  {10.1093/mnras/stw783}, \href
  {http://adsabs.harvard.edu/abs/2016MNRAS.459.1596S} {459, 1596}

\bibitem[\protect\citeauthoryear{{Steinberg}, {Coughlin}, {Stone}  \&
  {Metzger}}{{Steinberg} et~al.}{2019}]{2019MNRAS.485L.146S}
{Steinberg} E.,  {Coughlin} E.~R.,  {Stone} N.~C.,   {Metzger} B.~D.,  2019,
  \mn@doi [\mnras] {10.1093/mnrasl/slz048}, \href
  {https://ui.adsabs.harvard.edu/abs/2019MNRAS.485L.146S} {485, L146}

\bibitem[\protect\citeauthoryear{{Stone} \& {Metzger}}{{Stone} \&
  {Metzger}}{2016}]{2016MNRAS.455..859S}
{Stone} N.~C.,  {Metzger} B.~D.,  2016, \mn@doi [\mnras]
  {10.1093/mnras/stv2281}, \href
  {https://ui.adsabs.harvard.edu/abs/2016MNRAS.455..859S} {455, 859}

\bibitem[\protect\citeauthoryear{{Stone}, {Sari}  \& {Loeb}}{{Stone}
  et~al.}{2013}]{2013MNRAS.435.1809S}
{Stone} N.,  {Sari} R.,   {Loeb} A.,  2013, \mn@doi [\mnras]
  {10.1093/mnras/stt1270}, \href
  {https://ui.adsabs.harvard.edu/abs/2013MNRAS.435.1809S} {435, 1809}

\bibitem[\protect\citeauthoryear{{Stone}, {Kesden}, {Cheng}  \& {van
  Velzen}}{{Stone} et~al.}{2019}]{2019GReGr..51...30S}
{Stone} N.~C.,  {Kesden} M.,  {Cheng} R.~M.,   {van Velzen} S.,  2019, \mn@doi
  [General Relativity and Gravitation] {10.1007/s10714-019-2510-9}, \href
  {https://ui.adsabs.harvard.edu/abs/2019GReGr..51...30S} {51, 30}

\bibitem[\protect\citeauthoryear{{Stone}, {Vasiliev}, {Kesden}, {Rossi},
  {Perets}  \& {Amaro-Seoane}}{{Stone} et~al.}{2020}]{2020SSRv..216...35S}
{Stone} N.~C.,  {Vasiliev} E.,  {Kesden} M.,  {Rossi} E.~M.,  {Perets} H.~B.,
  {Amaro-Seoane} P.,  2020, \mn@doi [\ssr] {10.1007/s11214-020-00651-4}, \href
  {https://ui.adsabs.harvard.edu/abs/2020SSRv..216...35S} {216, 35}

\bibitem[\protect\citeauthoryear{{Turk}, {Smith}, {Oishi}, {Skory}, {Skillman},
  {Abel}  \& {Norman}}{{Turk} et~al.}{2011}]{yt}
{Turk} M.~J.,  {Smith} B.~D.,  {Oishi} J.~S.,  {Skory} S.,  {Skillman} S.~W.,
  {Abel} T.,   {Norman} M.~L.,  2011, \mn@doi [The Astrophysical Journal
  Supplement Series] {10.1088/0067-0049/192/1/9}, \href
  {http://adsabs.harvard.edu/abs/2011ApJS..192....9T} {192, 9}

\bibitem[\protect\citeauthoryear{{Ulmer}}{{Ulmer}}{1999}]{1999ApJ...514..180U}
{Ulmer} A.,  1999, \mn@doi [\apj] {10.1086/306909}, \href
  {https://ui.adsabs.harvard.edu/abs/1999ApJ...514..180U} {514, 180}

\bibitem[\protect\citeauthoryear{{Vandenbroucke} \& {De
  Rijcke}}{{Vandenbroucke} \& {De Rijcke}}{2016}]{2016A&C....16..109V}
{Vandenbroucke} B.,  {De Rijcke} S.,  2016, \mn@doi [Astronomy and Computing]
  {10.1016/j.ascom.2016.05.001}, \href
  {http://adsabs.harvard.edu/abs/2016A%26C....16..109V} {16, 109}

\bibitem[\protect\citeauthoryear{{Weissbein} \& {Sari}}{{Weissbein} \&
  {Sari}}{2017}]{2017MNRAS.468.1760W}
{Weissbein} A.,  {Sari} R.,  2017, \mn@doi [\mnras] {10.1093/mnras/stx485},
  \href {https://ui.adsabs.harvard.edu/abs/2017MNRAS.468.1760W} {468, 1760}

\bibitem[\protect\citeauthoryear{{Wu}, {Coughlin}  \& {Nixon}}{{Wu}
  et~al.}{2018}]{2018MNRAS.478.3016W}
{Wu} S.,  {Coughlin} E.~R.,   {Nixon} C.,  2018, \mn@doi [\mnras]
  {10.1093/mnras/sty971}, \href
  {https://ui.adsabs.harvard.edu/abs/2018MNRAS.478.3016W} {478, 3016}

\bibitem[\protect\citeauthoryear{{Yalinewich}, {Steinberg}  \&
  {Sari}}{{Yalinewich} et~al.}{2015}]{2015ApJS..216...35Y}
{Yalinewich} A.,  {Steinberg} E.,   {Sari} R.,  2015, \mn@doi [\apjs]
  {10.1088/0067-0049/216/2/35}, \href
  {http://adsabs.harvard.edu/abs/2015ApJS..216...35Y} {216, 35}

\bibitem[\protect\citeauthoryear{{Yalinewich}, {Guillochon}, {Sari}  \&
  {Loeb}}{{Yalinewich} et~al.}{2019}]{2019MNRAS.482.2872Y}
{Yalinewich} A.,  {Guillochon} J.,  {Sari} R.,   {Loeb} A.,  2019, \mn@doi
  [\mnras] {10.1093/mnras/sty2809}, \href
  {https://ui.adsabs.harvard.edu/abs/2019MNRAS.482.2872Y} {482, 2872}

\bibitem[\protect\citeauthoryear{{van Velzen}}{{van
  Velzen}}{2018}]{2018ApJ...852...72V}
{van Velzen} S.,  2018, \mn@doi [\apj] {10.3847/1538-4357/aa998e}, \href
  {https://ui.adsabs.harvard.edu/abs/2018ApJ...852...72V} {852, 72}

\makeatother
\end{thebibliography}

\bsp	
\label{lastpage}
\end{document}